# Mechanical Research and Development of monocrystalline silicon neutron beam window for CSNS

Zhou Liang (周良)[1,2;1]  Qu Hua-Min (屈化民)[1]

[1] Institute of High Energy Physics, Chinese Academy of Sciences, Beijing 100049, China

[2] University of Chinese Academy of Sciences, Beijing 100049, China

**Abstract:** The monocrystalline silicon neutron beam window is one of the key components of neutron spectrometers and thin circular plate. Monocrystalline silicon is a brittle material and its strength is not constant but is consistent with the Weibull distribution. The window is designed not simply through the average strength, but according to the survival rate. Bending deformation is the main form of the window, so dangerous parts of the neutron beam window is stress-linearized to the combination of membrane stress and bending stress. According to the Weibull distribution of bending strength of monocrystalline silicon based on a large number of experimental data, finally the optimized neutron beam window is 1.5mm thick. Its survival rate is 0.9994 and its transmittance is 0.98447; it meets both physical requirements and the mechanical strength.

## 1 Introduction

For China Spallation Neutron Source (CSNS), neutrons move through every place either in a vacuum environment, or in special atmosphere, such as helium. Almost neutron spectrometers have neutron beam windows that are indoors or outdoors for neutrons [1]. Neutron beam window is one of the key components; it can make neutrons move through with enough loss and little neutron background. The international neutron spectrometers generally use three kinds of materials for beam windows: monocrystalline silicon, sapphire and aluminum alloy. From the physical point of view, silicon is the best material and neutron beam window is as thin as possible; but silicon is a brittle material and neutron beam window also meets enough mechanical strength to withstand the pressure, so it is as thick as possible. The development of neutron beam windows that meet the physical requirements and satisfy the mechanical strength is the urgent need for all neutron spectrometers.

## 2 Problems in the development

The prototype neutron beam window is made from monocrystalline silicon and has sizes: 40mm in diameter and 0.5mm in thickness. The basic mechanical properties of monocrystalline silicon [2] are showed in tabel1.

Table 1  the basic mechanical properties of monocrystalline silicon

| Features | Parameters |
| --- | --- |
| Density (293K) | 2.327g/cm3 |
| Lattice constant (293K) | 541.962pm |
| Knoop hardness | 9.5-11.5GPa |
| Elastic coefficient | $C_{11}$=1.6564e11Pa |
|  | $C_{12}$=0.6394e11Pa |
|  | $C_{44}$=0.79514e11Pa |
| Shear strength | 240MPa |
| Tensile strength | 350MPa |
| Compressive strength | 950MPa |
| Bending strength | 300-1000 MPa |

---
[1)] Email: liangzhou@ihep.ac.cn

| Young's modulus | $E_{[100]}$=1.30e5MPa |
| --- | --- |
| | $E_{[110]}$=1.68e5MPa |
| | $E_{[111]}$=1.87e5MPa |
| Poisson's ratio | $P_{[100]}$=0.278 |
| | $P_{[110]}$=0.25 |
| | $P_{[111]}$=0.20 |

As the same as the design of LENS(As shown in Figure 1), there is a rubber sealing ring in each surface and so silicon window is not direct contact with the metal flange.

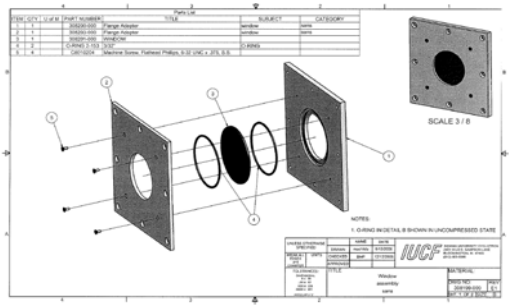

Figure 1  The neutron beam window of LENS

As shown in Figure 2, from the finite element analysis

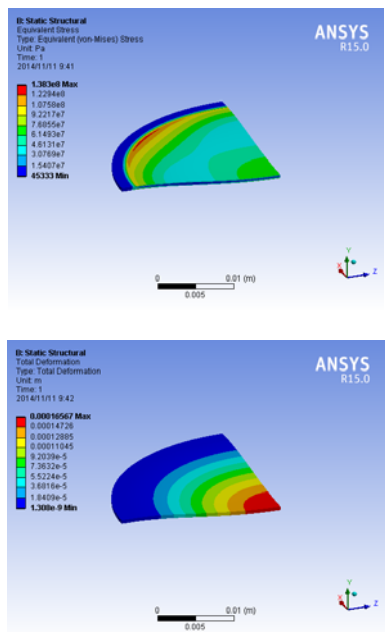

Figure 2  the finite element analysis results of monocrystalline silicon neutron beam window（0.5mm thick） under an atmospheric stress results the maximum equivalent stress is 138.3MPa that does not exceed the tensile strength (350MPa) when the window withstand an atmospheric pressure, and the safety coefficient is 2.53, the displacement is less than 0.166mm. But the window is still broken during the process of vacuum extraction, as shown in Figure 3.

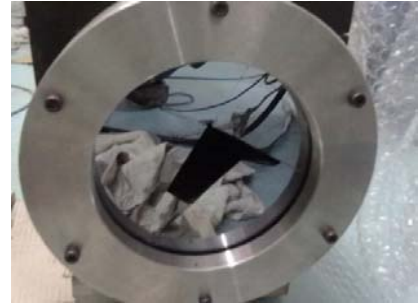

Figure 3  the broken monocrystalline silicon neutron beam window

## 3      The cause of failure

Monocrystalline silicon is a brittle material and the neutron beam window is thin circular plate. Its bending strength is not a constant value and is consistent with some statistic distribution, generally the Weibull distribution is used to describe the bending strength of brittle materials [3][5]. Under some bending stress $\sigma_b$ the window's failure rate is: $F(x) = 1 - \exp\left[-\left(\frac{\sigma_b}{a}\right)^m\right]$, and the survival rate is: $P = \exp\left[-\left(\frac{\sigma_b}{a}\right)^m\right]$, I.e. $\ln\ln\frac{1}{p} = m\ln\sigma_b - m\ln a$ or $y = mx - b$.

$m$ is the heterogeneity, and used to measure the material inhomogeneity, the greater the value, the more homogeneous. Simply supported wafer is applied with concentrated load impact to detect bending strength of monocrystalline silicon in GB/T 15615-1995. It supports enough dates based on a large number of experimental

Table 2  bending strength of different thick monocrystalline silicon

| | Thickness(mm) | | | | | | The trend value |
|---|---|---|---|---|---|---|---|
| | 0.25 | 0.3 | 0.5 | 0.6 | 0.75 | 0.9 | |
| bending strength (MPa) | 132.4 | 135.3 | 147.1 | 147.1 | 147.1 | 147.1 | 147.1 |
| | 160.8 | 164.8 | 180.4 | 186.3 | 196.1 | 196.1 | 196.1 |
| | 183.9 | 193.7 | 215.8 | 223.1 | 237.8 | 245.2 | 245.2 |
| | 211.8 | 217.7 | 250.1 | 261.8 | 276.6 | 294.2 | 294.2 |
| | 236.8 | 243.7 | 278.0 | 295.2 | 319.2 | 339.8 | 343.2 |
| | 258.9 | 270.7 | 309.9 | 329.5 | 357.0 | 380.5 | 392.3 |
| | 286.9 | 295.7 | 339.8 | 357.5 | 392.8 | 423.7 | 441.3 |
| | 313.8 | 323.6 | 367.8 | 392.3 | 426.6 | 460.9 | 490.3 |
| | 334.4 | 350.6 | 393.7 | 420.7 | 458.5 | 501.6 | 539.4 |
| | 364.8 | 376.6 | 423.7 | 453.1 | 494.3 | 535.4 | 588.4 |
| | 395.2 | 401.6 | 452.6 | 478.1 | 522.7 | 573.7 | 637.4 |
| | 418.8 | 432.5 | 480.5 | 508.0 | 556.0 | | 686.5 |
| | 448.7 | 463.4 | 507.5 | 536.9 | | | 735.5 |
| | 478.6 | 494.3 | 533.5 | 564.9 | | | 784.5 |
| | 508.5 | 525.2 | 566.8 | 591.8 | | | 833.6 |
| | 538.4 | 556.0 | 600.2 | 626.7 | | | 882.6 |
| | 568.3 | 586.9 | | | | | 931.6 |
| | 597.8 | 617.4 | | | | | 980.1 |

data for detecting bending strength of different thick silicon [4], as showed in table 2.

Fitting the data from table 2, When the thickness is 0.25mm, $y = 2.3705x - 14.246$, when the thickness is 0.30mm, $y = 2.3705x - 14.323$, when the thickness is 0.5mm, $y = 2.5171x - 15.263$, when the thickness is 0.6mm, $y = 2.4505x - 14.994$, when the thickness is 0.75mm, $y = 2.453x - 14.802$, when the thickness is 0.9mm, $y = 2.3065x - 13.974$. If in accordance with the trend value, $y = 1.9138x - 12.397$. Thus the homogeneity degree of 1.9138 is the minimum value of silicon materials. It is the most reliable if the window is design according to the extreme situation of the minimum homogeneity degree, and the survival rate is:

$$P = \exp\left[-\left(\frac{\sigma_b}{650.44}\right)^{1.9138}\right]$$

Because the silicon window is a thin circular plate, bending deformation is the main form; it is equivalent to the combination of membrane stress and bending stress [6], then dangerous parts of the neutron beam window is stress-linearized, the result is showed in fatigue 4. From the result, the maximum bending stress is about 115 MPa. And so the survival rate is 0.9644. It is that there is one broken when thirty beam windows are used. Therefore, it is very possible that the prototype neutron beam window was broken. And the window needs to be thicker.

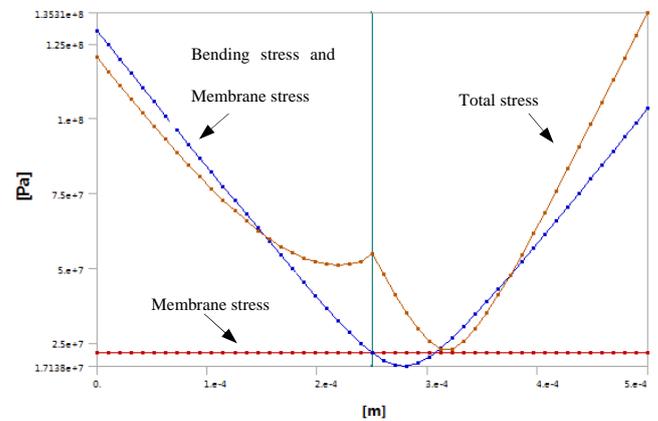

Figure 4  the result of stress-linearization of monocrystalline silicon neutron beam window（0.5mm thick）

## 4 Optimization of the thickness of monocrystalline silicon neutron beam window

As shown in fatigue 5, when the silicon window thickness increased to 1mm, the maximum equivalent stress is 38.9MPa; the maximum displacement is 22 micron.

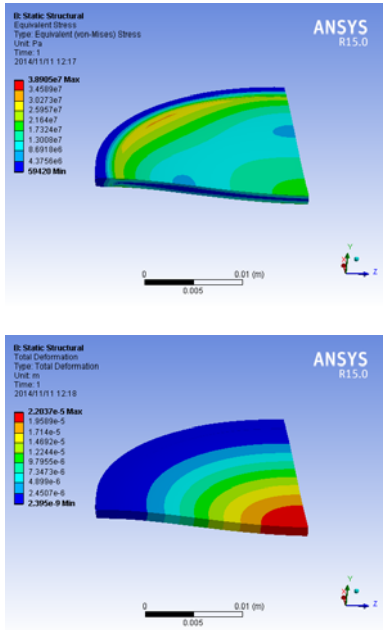

Figure 5    the finite element analysis results of monocrystalline silicon neutron beam window（1mm thick）under an atmospheric stress

Then dangerous parts of the neutron beam window is stress-linearized, the result is showed in fatigue 6. From the result, the maximum bending stress is about 30MPa. And so the survival rate is 0.9972. It is that there is about one broken when three hundred beam windows are used.

The mean value of the window's bending strength is:

$$E(\sigma_b) = \sigma_{b0}\Gamma(1+\frac{1}{m}) = 650.44 \times \Gamma(1+\frac{1}{1.9138}) \approx 577.04 MPa$$

And the mean variance of the window's bending strength is:

$$D(\sigma_b) = \sigma_{b0}\sqrt{\left[\Gamma(1+\frac{2}{m})-\Gamma^2(1+\frac{1}{m})\right]} = 650.44 \times \sqrt{\left[\Gamma(1+\frac{2}{1.9138})-\Gamma^2(1+\frac{1}{1.9138})\right]} \approx 300.31 MPa$$

Therefore, if the design in accordance with the mean value of the window's bending strength, the safety coefficient is: $\lambda = \frac{E(\sigma_b)}{30} \approx 19.24$.

If the window is replaced easily and conveniently, 1mm thick window is enough to withstand stress.

As shown in fatigue 7, when the silicon window thickness increased to 1.5mm, the maximum equivalent stress is 20.2MPa; the maximum displacement is 7 micron.

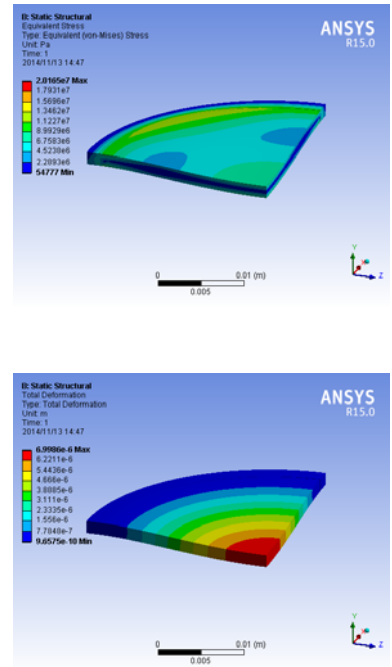

Figure 7    the finite element analysis results of monocrystalline silicon neutron beam window（1.5mm thick）under an atmospheric stress

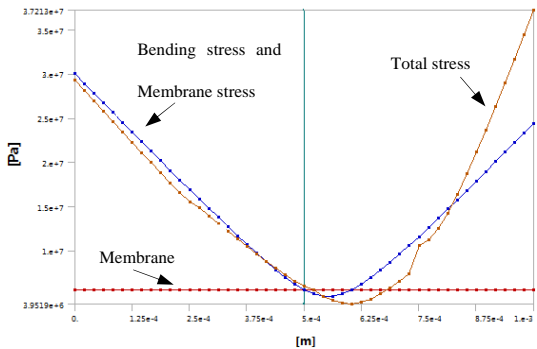

Figure 6    the result of stress-linearization of monocrystalline silicon neutron beam window（1mm thick）

Then dangerous parts of the neutron beam window is stress-linearized, the result is showed in fatigue 8.

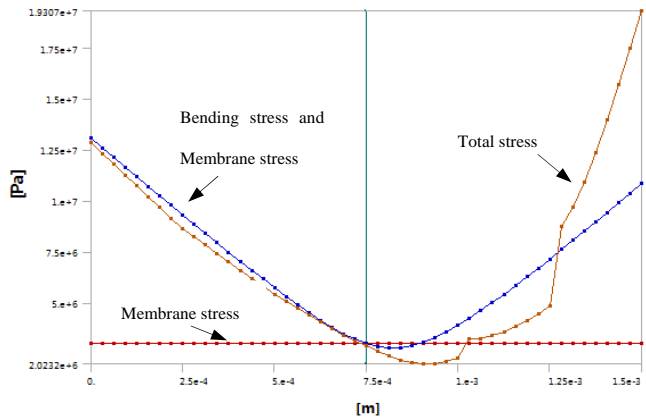

Figure 8 the result of stress-linearization of monocrystalline silicon neutron beam window（1.5mm thick）

From the result, the maximum bending stress is about 13.1MPa. And so the survival rate is 0.9994. It is that there is one broken when sixteen hundred sixty-seven beam windows are used. If the window is replaced uneasily and inconveniently, less than 1.5mm thick window is enough to withstand stress. Therefore, if the design in accordance with the mean value of the window's bending strength, the safety coefficient is:

$$\lambda = \frac{E(\sigma_b)}{13.1} \approx 44.05$$.

The transmittance of 1.5mm thick silicon neutron beam window is 0.98447. It meets the physical requirements. And 10 neutron beam windows tested are not broken. It satisfies the mechanical strength.

## 5 Conclusion

The monocrystalline silicon neutron beam window is thin circular plate. According to the Weibull distribution of bending strength of monocrystalline silicon, finally the optimized neutron beam window is 1.5mm thick and its survival rate is 0.9994; it meets both physical requirements and the mechanical strength.

Discreteness of the mechanical strength need be considered when using brick materials .The design way provides a reference for large size sapphire neutron beam window in future.

―――――――――――――――――――――――――

## References

1. WANG FangWei，LIANG TianJiao，YIN Wen，YU QuanZhi，HE LunHua，TAO JuZhou，ZHU Tao，JIA XueJun，ZHANG ShaoYing. Physical design of target station and neutron instruments for China Spallation Neutron Source [J]. SCIENCE CHINA: Physics, Mechanics & Astronomy, 2013,12:2410-2424.

2. Sundararajan S, Bhushan B. Fracture Development of AFM-based Techniques to Measure Mechanical Properties of Nanoscale Structures [J]. Sensors and Actuator s A: Physical, 2002, 101(3): 33 8-351.

3. JIN Zong-Zhe, BAO Yi-Wang, Characterization of Mechanical Properties for Brittle Materials and Ceramics [M], Beijing: Chinese Railway Publishing House, 1996.

4. GB/T 15615-1995: Test method for measuring flexure strength of silicon slices.

5. JIN Zong-Zhe, Ma Juan-Rong The optimum numbers of samples estimated by Weibull shape parameter in the statistic analysis of strength of brittle materials [J]. Journal of the Chinese Ceramic Society, 1989, 17(3): 229-236.

6. Wang Xin-Ming, W.Z. Mike. Design and Calculation of Engineering Pressure Vessel [M], Beijing: National Defense Industry Press, 2011.10.